\documentclass[iop]{emulateapj-rtx4} % read by: ACROREAD
\shortauthors{Sekanina}
\shorttitle{Deceleration of Encke's Comet}
\slugcomment{Version \today }

\begin{document}
\title{THE END OF AN ERA IN COMETARY ASTRONOMY:\\
       THE DECELERATION OF COMET ENCKE}
\author{Zdenek Sekanina}
\affil{Jet Propulsion Laboratory, California Institute of Technology,
  4800 Oak Grove Drive, Pasadena, CA 91109, U.S.A.}
\email{Zdenek.Sekanina@jpl.nasa.gov.}

\begin{abstract} % maximum length = 1920 characters; estimate = 1933 (1919)
It is noted that effective very recently, the orbital motion of Encke's
comet has become affected by a very slight nongravitational
deceleration.  Soon after J.\,F.\,Encke established in the early 19th
century that the comet was returning to perihelion every 3.3~years, he
also discovered that the object was notorious  
% deceleration.  Ever since J.\,F.\,Encke's discovery in the early 19th
% century of the failure of the comet to obey the gravitational law, it was
% notorious
for returning to perihelion a little earlier than predicted by the Newtonian
theory.  The acceleration persisted over a period of two centuries, but
its rate was gradually decreasing.
Generations of cometary astronomers were curious to know whether or not the
comet would eventually move in purely gravitational orbit.  A model
based on the assumption of a precession of the comet's nucleus, which
predicted that the acceleration would change into a deceleration, was
not published until 1979.  This transition has now been documented
by two independent, highly-accurate orbit determinations.  The era
of the comet's persevering nongravitational acceleration is finally over.  
\end{abstract}

\section{Brief History of the Orbital Motion of Comet Encke} %%% Sec. 1
For several generations any student of astronomy learnt two basic
facts about comets:\ Halley's was the first to provide evidence
that some comets move in closed orbits, returning to perihelion
periodically at predictable times; Encke's was the first to provide
evidence that some periodic comets do not return as predicted by
the Newtonian theory, being subjected to nongravitational
perturbations.  The significance of Encke's acceleration was
enormous, prompting, almost exactly two centuries ago, the
introduction of a hypothesis of resisting interplanetary medium.
And while this conjecture was later discredited, the acceleration
(sometimes referred to as the secular acceleration) of the comet
remained undisputed scientific fact.  Studies of the comet, which
has been returning to perihelion every 3.3~years and was observed at
more than 60~returns (and for decades now along the entire orbit),
continued to confirm the anomaly over and over again, even
though with one important proviso:\ starting after the mid-19th century,
the {\it rate of acceleration\/} was noticed to be gradually
decreasing with time.  And then the orbital anomaly
began to display another peculiarity:\ the {\it rate of diminishing\/}
became the slower the lower was the acceleration, as if the effect were converging
to nil.  This trend kept continuing over agonizingly long periods
of time until very recently.

In the 1960s, there still was no consensus on the issue of what
happens after the acceleration vanishes, if it does so at all:\ would it
turn into a deceleration or would the comet's motion perpetually obey
the gravitational law?  In fact, there was a third, less likely
possibility:\ the acceleration may reach a minimum and then start
increasing again.  Only in the 1970s, as the research on the
rotation and precession of cometary nuclei began to pick up
steam, was it established that there was nothing sacred about the
acceleration and that its turning into a deceleration was inevitable
and only a matter of time.

\section{Long-Term Variations in Comet Encke's Nongravitational
 Acceleration} %%% Sec. 2
The history of investigation of Encke's orbital motion makes a
fascinating story.  I described it in detail elsewhere
(Sekanina 1991).  A few highlights follow.

\subsection{Work by J.\,F.\,Encke} %%% Sec. 2.1
Encke (1819, 1820) discovered that --- over the period of
1786--1819 or 10~revolutions about the Sun --- the comet had been
arriving at perihelion systematically {\it earlier\/} than predicted
after the planetary perturbations were accounted for; its net orbital
period was getting progressively shorter.  The differences averaged
about 2.7~hours per revolution per each revolution, accumulating to
a respectable difference of 5~days over the 10~cycles.

Encke (1823) soon came up with his infamous hypothesis of a resisting
interplanetary medium to explain the orbital anomaly.  He assumed that
the spatial density of the medium varied inversely as the square of
heliocentric distance $r$ and that the force of resistance, $U$, exerted on
the comet, varied as the medium's density and as the square of the
comet's orbital velocity $V$, \mbox{$U = U_0 V^2/r^2$}, where $U_0$
was a constant to be determined from the observations.  Encke
concluded that the force affected the comet's mean motion (and
therefore orbital period) and eccentricity, but not the positions
of the line of apsides and the orbital plane.  
 
Encke spent a significant fraction of his lifelong scientific activity
on the comet.  The 1858 return was the last that he worked on (Encke
1859), finding that the acceleration remained essentially constant over
the period of seven decades.  At various times he measured its magnitude
either by the constant $U_0$ of the force, or by the change in the
orbital period, $\Delta P$, or by the change in the mean daily motion,
$\Delta \mu$.  Over time, the parameter used the most became known
as the {\it secular acceleration coefficient\/} $\kappa$, related to
$\Delta P$ by
\begin{equation}
\kappa = -73^{\prime\prime\!}.92 \, \frac{\Delta P}{P},
\end{equation}
where $\Delta P$ is in hours per revolution per revolution and $P$ in
years.  For Encke's comet \mbox{$P = 3.30$ yr} and the secular
acceleration coefficient becomes
\begin{equation}
\kappa = -22^{\prime\prime\!}.4 \: \Delta P.
\end{equation}
For the apparitions investigated by Encke, 1786--1858, \mbox{$\Delta P
\simeq -2.7$ hr} and therefore \mbox{$\kappa \simeq 60^{\prime\prime}$}.

The invariability of the anomaly's magnitude was very reassuring to
Encke, and it is no wonder that~until~his death in 1865 he
believed that the resisting-medium hypothesis was valid.

As will become apparent from the following, the apparent invariability
of the resisting medium was a result of unusual coincidence.  The
magnitude of the comet's acceleration was already {\it diminishing\/}
from the 1820s on, but previously it had been {\it increasing\/} in
a symmetric fashion between 1786 and 1820.  The peak was flat and
only about 10~percent above the minimum at either end of the
seven-decades long period.  If Encke were given an opportunity to
include a few returns following 1858, he would have been greatly
disappointed.  

\subsection{Work by E.\,von Asten, O.\,Backlund, and\\the Russian
 School} %%% Sec. 2.2
After Encke's death, further research on the comet's motion in Berlin
was soon discontinued.  The focus of activity moved to the Pulkovo
Observatory near Saint Petersburg, after von Asten left Berlin to
join the Observatory's staff.  His attempts to link the apparitions
1819--1871 failed miserably.  And then came Backlund, an import from
Sweden, who took over after von Asten's untimely death in 1878.

Backlund's extensive computations indicated that the failure of von
Asten's work was owing to a sharp decrease in the comet's acceleration.
Subsequently he met the same problem as he tried to link additional
returns up to 1908.  Contrary to Encke's value of 60$^{\prime\prime}$,
Backlund found that $\kappa$ equaled 50$^{\prime\prime}$ for the
returns 1858--1868, 42$^{\prime\prime}$ for 1868--1895, and barely
one half Encke's value for 1895--1905.  With no cures working, he
finally was compelled to admit that ``[t]he cause of the acceleration
of the mean motion cannot be a resisting medium of so simple a nature
as Encke supposed \ldots'' (Backlund 1910).

Backlund's work was continued by a number of Soviet astronomers, using
essentially the same technique of analysis as Encke and Backlund.
With the hypothesis of resisting medium abandoned, but no new model
firmly in place, the behavior of the mechanism exerting the acceleration
was entirely unclear.  It is therefore not surprising that Makover
(1955), the most influential member of the new Russian School,
introduced a major simplification by assuming that the entire effect
was produced as an impulse at perihelion.  The effort of the group was
focused on gradually extending the database in the hope of getting a
more profound insight into the matter.

\subsection{From Hypothesis of Resisting Medium\\to Whipple's Model}
%%% Sec. 2.3

Flawed paradigms are a routine --- and, to a degree, useful --- part of
the process of searching for the correct solution to any problem.  The
unfortunate circumstance about the hypothesis of resisting medium was not
that it was wrong, but that too long was it taken for granted that it
was correct.  It was not challenged even when it began to display major
cracks, perhaps because of Encke's reputation as a brilliant scientist.
Among the few astronomers criticizing the hypothesis, the most vocal was
Bessel (1836a, 1836b), who objected to it on the grounds that there
was no other evidence for the existence of a resisting medium and that
the motions of comets that were decelerated could not be explained.  Instead,
Bessel argued that a recoil effect by the mass outflowing from the nucleus
on its motion, the orbital period in particular, was inevitable.
Unfortunately, Bessel's far-sighted ideas, which he formulated in
reaction to his own physical observations of Halley's comet, were
ignored by his peers and later largely forgotten.  Only Bessel's
investigations on cometary tails were continued by T.\,Bredikhin
(Jaegermann 1903).

More than a century after Bessel's pioneering work, Encke's comet
and its acceleration once again took center stage, appearing in
the title of one of the most celebrated comet papers of all time
(``A Comet Model. I. The Acceleration of Comet Encke''), in which
Whipple (1950) introduced his influential icy-conglomerate model,
interpreting the acceleration as a recoil effect by outgassing,
primarily of water ice, from a rotating nucleus and used the
acceleration's magnitude to estimate the comet's relative
mass-loss rate.  Sublimation lags were used to explain the
sign of the transverse component of the nongravitational effect in
terms of the rotation sense:\ an acceleration indicated a retrograde
rotation, a deceleration a prograde rotation.  As I noted in Section~1,
a decline in the magnitude of the nongravitational effect was
typically interpreted as a drop in the rate of ejected mass, thus
potentially implying an aging comet that is gradually turning into
an asteroid.

Whipple's work set in motion vigorous activity on related
issues, directed at gaining a greater insight into the problem of
nongravitational effects.  This activity included a major effort by
Marsden (1969, 1970) aimed at developing a new orbit-determination code
with nongravitational terms incorporated directly into the equations of
motion, thus allowing the derivation of the parameters for
the components of the nongravitatonal force in an orthogonal system of
coordinates.  Subsequently, in the so-called Style~II version of the
code its initial weakness was removed by incorporating a realistic
nongravitational law, modeled by the sublimation curve of water ice from
a rapidly rotating spherical nucleus (Marsden et al.\ 1973).  The functional
dependence on heliocentric distance is flexible enough that by changing
a single parameter it is possible to accommodate outgassing models for 
other ices or (near the Sun) refractory materials.

\begin{table*}
\vspace{-4.15cm}
\hspace{0.55cm}
\centerline{
\scalebox{1}{
\includegraphics{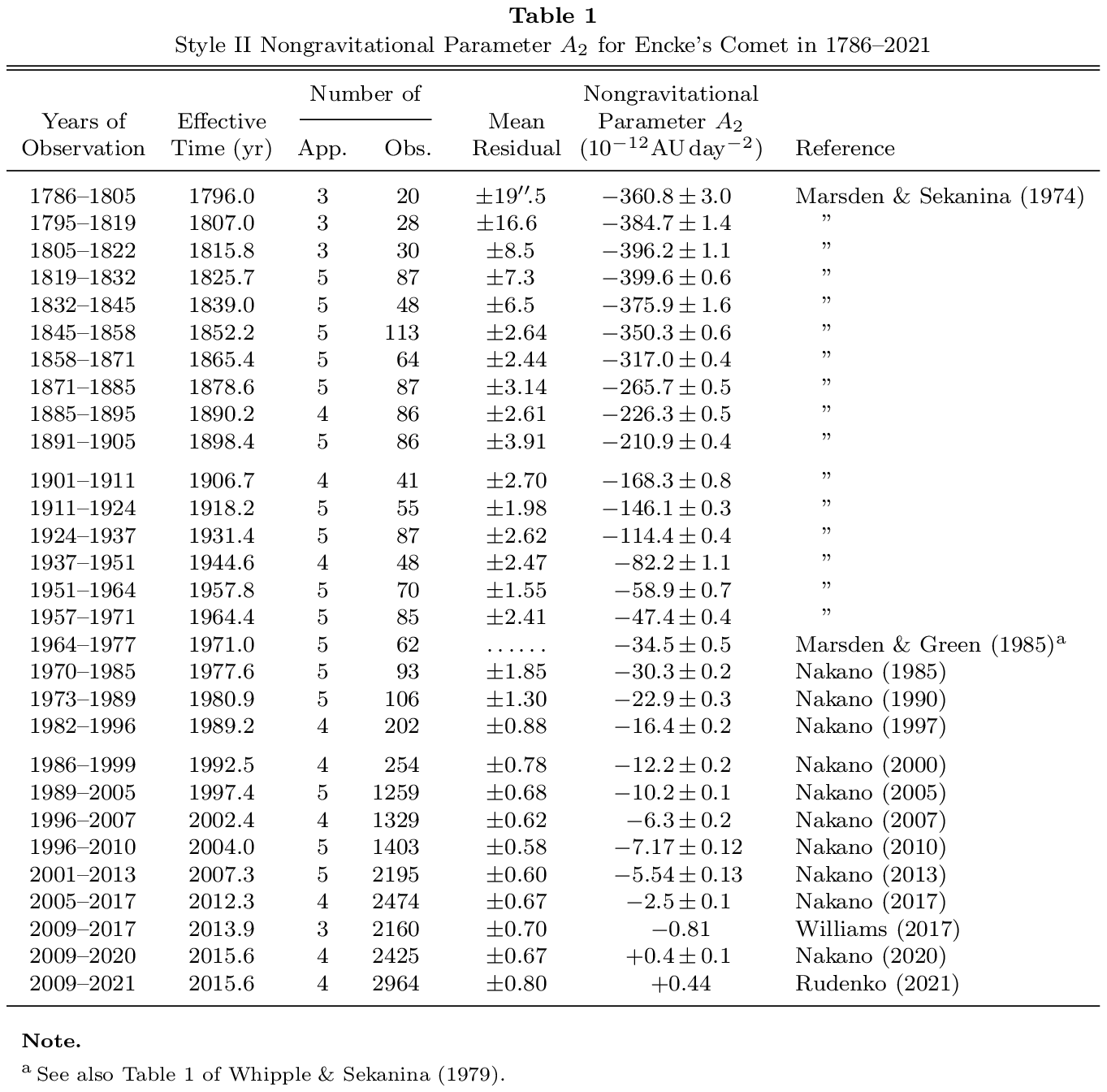}}}
\vspace{-11.2cm}
\end{table*}

\subsection{The End of an Era:\ After 230 Years\\Encke's Comet Is Decelerated}
%%% Sec. 2.4
%
Equipped with the Style II nongravitational orbit determination code, the
task of complete characterization of the motion of Encke's comet was reduced
to painstakingly collecting the astrometric observations dated back to
1786.  Marsden had already gathered critical photographic observations,
which he strongly preferred to visual ones (the early 1970s were of course
a pre-CCD era).  However, there were hardly any photographic observations
made before the 1920s, so much work was still ahead of us.  As expected, the
farther into the past we got, the more time consuming the reduction of the
astrometric data became and the more inaccurate the reduced positions of the
comet turned out to be.

By virtue of being a function of heliocentric distance, the nongravitational
law is symmetric relative to perihelion.  It is straightforward to show
that when linking a number of apparitions with such a law, the radial component
of the nongravitational perturbation, described by the parameter $A_1$,
actually measures the effect on the argument of perihelion (Sekanina 1993).
$A_1$ is often poorly determined and of little interest.  Important is the
transverse component, whose parameter $A_2$ describes the nongravitational
perturbation of the mean motion and is closely related to the secular
coefficient $\kappa$ (Section~2.1).

Our work on the motion of Encke's comet (Marsden \& Sekanina 1974)
terminated with the return of 1971, but it was more recently continued
(with the same Style~II technique) by others, S.\,Nakano in particular.
A representative set of orbital runs is presented in self-explanatory
Table 1 and is plotted in Figure~1.  To accentuate the transition from
the acceleration to {\vspace{-0.04cm}}deceleration,~I~plot~the data
linearly in $\sqrt{|A_2|}\,{\rm sign}(A_2)$.  A linear $A_2$ scale shows the
effect clearly only when greatly blown up, as in the inset of Figure~1
covering the years 2000-2020.

\section{Rotation, Active Areas, and\\Coma Morphology} %%% Sec. 3
In the 1970s the perception of comets began to change dramatically.  The
morphology of the coma's appearance became the subject of interest:\ why
do comets display narrow jets, concentric halos, spirals, or,
as does Encke's comet before perihelion, a conic-shaped fan directed
generally towards the Sun?  A consensus began to grow that such features
of anisotropic outgassing could be explained only if (a)~they originate from
particular, active areas on the surface of the cometary nucleus and (b)~the
nucleus rotates.  If so, the observed morphology could be used to learn
about the locations of the active sources and about the comet's rotation
vector, that is, the spin rate and the orientation of the rotation axis.

\begin{figure*}
\vspace{-5.9cm}
\hspace{-0.2cm}
\centerline{
\scalebox{0.86}{
\includegraphics{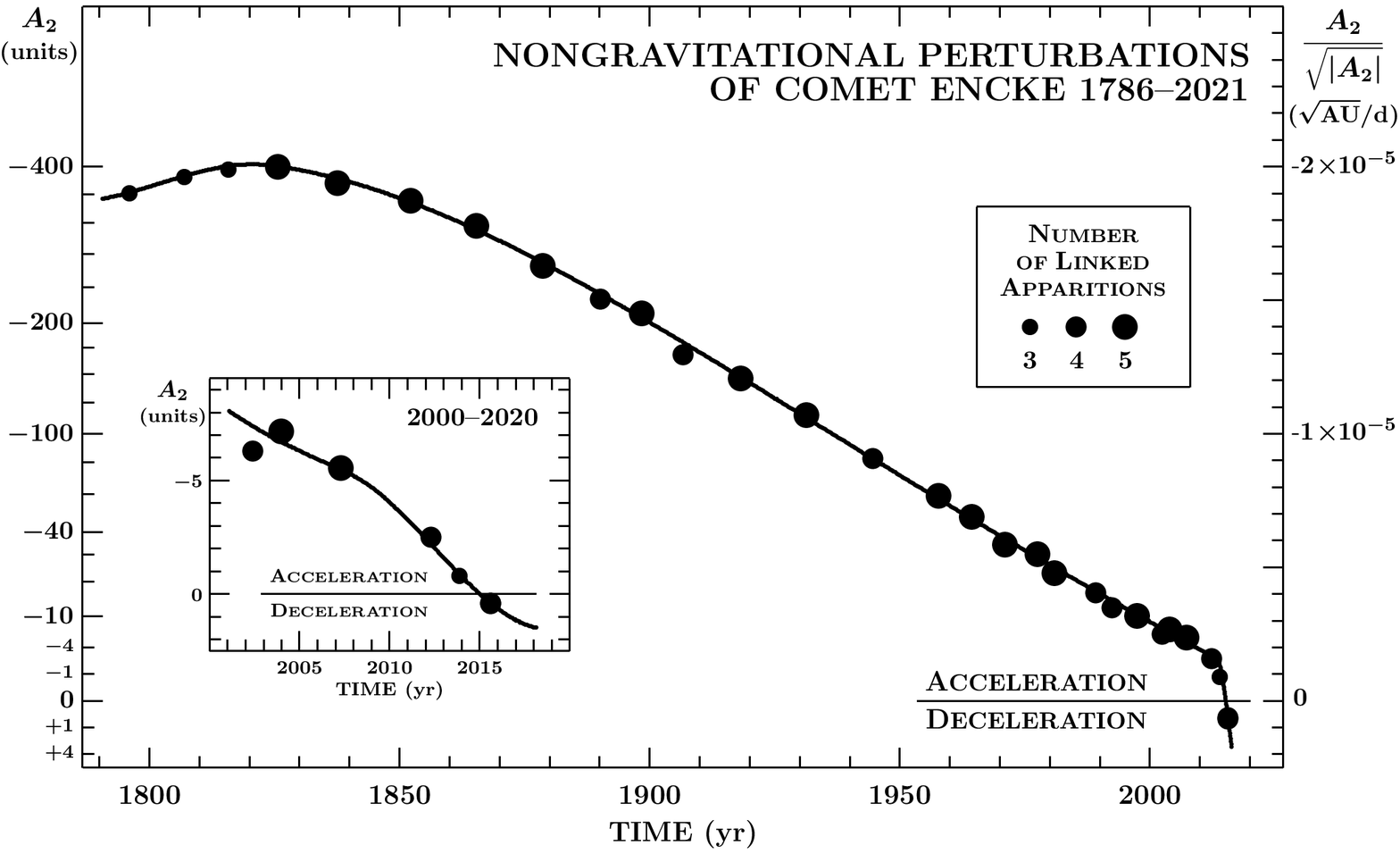}}}
\vspace{-8.65cm}
\caption{Plot of the parameter $A_2$ of the transverse component of the
{\vspace{-0.04cm}}nongravitational perturbation of the nucleus of Encke's
comet over the entire period of observation.  To emphasize the recent
transition from the {\vspace{-0.05cm}}acceleration to deceleration, the
plot is linear in $\sqrt{|A_2|}$ with the $A_2$ sign preserved (right
scale).  For comparison, the nonlinear scale for $A_2$ is on the left
(1~unit = 10$^{-12}$\,AU day$^{-2}$).  Table~1 shows that the magnitude of
the deceleration derived from the runs 2009--2020/21 is only 0.1~percent
of the peak acceleration in the 1820s, as shown in the inset.  The
equivalent $\Delta P$ is +10~seconds compared to the peak change of
$-$2.7~hours per revolution per revolution.{\vspace{0.8cm}}}
\end{figure*}

As for Encke's comet, a sunward oriented fan-shaped coma exhibited
upon approach to perihelion was a standard morphological feature observed
return after return.  I began my work by examining the pattern of deviations
from the sunward direction in the orientation of the fan for four periodic
comets, including Encke.  I found that the established sense of rotation
correlated with the results derived from the transverse component of the
nongravitational force (Sekanina 1979).  The implication for Encke's comet
was that the temporal variations in its nongravitational acceleration could
be related to a changing position of the comet's polar axis as a result of
torques exerted by the anisotropic outgassing from a discrete active center
(or centers) on its surface.  A precession model developed by Whipple \&
Sekanina (1979) showed that the observed curve of nongravitational
acceleration, shown here in Figure~1, was consistent with such a scenario.

The precession model was necessarily predicated on uncertain assumptions
about the rotation period, the nuclear size, the water production law
(approximated by the light curve), etc.\footnote{Also, at the time I still
struggled with the relationship between the fan-shaped coma orientation and the
projected position of the rotation axis, which was an important part of the
precession model.  The employed technique to determine the position of the
rotation axis was preliminary.  It was not until the late 1980s that the
final version of the relationship was completed (Sekanina 1987) and applied
to Encke's comet (Sekanina 1988a, 1988b).}  Yet the plausibility of the nuclear
precession as the trigger for the nongravitational perturbations of the
orbital motion of Encke's comet could hardly be disputed.  It was this 
precession model that predicted the inevitability of the change from the
acceleration to deceleration, even though on account of the many approximations
the time of this historic transition could not be predicted with high accuracy.
 
\section{The Light Curve}
The light curve of Encke's comet used to be the subject of disputes about
the rate at which the comet has been fading (e.g., Meisel 1969 and the
references therein), the estimates varying from 1 mag to 3 mag per century.
Although the two numbers were so far apart that they could not be both correct,
Table~2 shows that the presence of short-term fluctuations superposed on
the systematic trend complicated the matter.  The table lists the absolute
visual magnitude $H_0$ (reduced to unit heliocentric and geocentric
distances) of Encke's comet at five apparitions over a period of 33~years,
derived by Beyer (1972), one of the most experienced visual comet observers ever,
from his own photometric observations.  The average rate of fading amounts
to \mbox{$2.3 \!\pm\! 1.1$ mag} per century.  The numbers show that the
rate of decline is practically nil between 1937 and 1961, but nominally more
than 6~mag per century between 1961 and 1971.

\begin{table}[t]
\vspace{-11.02cm}
\hspace{-0.45cm}
\centerline{
\scalebox{1}{
\includegraphics{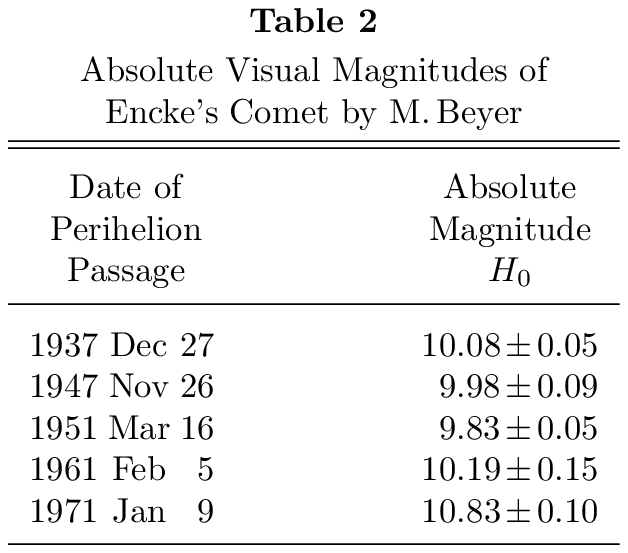}}}
\vspace{-12.1cm}
\end{table}

The problem appears to have been settled by Kam\'el (1991), who concluded that
Encke's comet suffered no secular fading, but that the strong perihelion
asymmetry of the light curve in the 1840s, when it peaked three weeks before
perihelion, vanished by the late 20th century.  Rather than moving the light
curves from various perihelion returns up and down to check the fading, Kam\'el
moved them left and right to check the asymmetry.  The equivalence between the
rates in the two directions in terms of the absolute magnitude can easily be
estimated; given that Encke's brightness normalized to 1~AU from the Earth
varies with heliocentric distance $r$ approximately as $r^{-6}$ near 1~AU
from the Sun and that between 1~AU and perihelion the comet's heliocentric
distance changes at an average rate of 0.016~AU per day, a decline of $\Delta
H_0$ in the absolute magnitude is equivalent to a forward shift of the light
curve's peak by \mbox{$\Delta t_{\rm peak} = 10 \, \Delta H_0$ days}.  Kam\'el's
suggestion that the peak shifted by about 25~days over 150~years is thus
equivalent to $\Delta H$ of 1.7~mag per century.

Most importantly, Kam\'el did point out that the disappearance of the light
curve's perihelion asymmetry appears to have correlated with the disappearance
of the nongravitational acceleration in the comet's orbital motion, a further
indication that the dynamical effect was linked to the comet's activity.

\section{Concluding Remarks}
This paper presents my recollection of an era in the life of Encke's comet.
The nongravitational acceleration of the comet's motion was deemed
scientifically so important that 130~years after its discovery it
made it into the title of the paper presenting the most successful comet model.
In the hope that from now on the comet's deceleration will be equally firmly
established in the course of this century, I have given this contribution
a headline that paraphrases the title of Whipple's famous paper.

The history of investigation of the unusual motion of Encke's comet is rather
typical of a treatment that a major discovery is subjected to.  Seldom is the
first hypothesis that enjoys broad initial support found eventually to be the
correct one.  The unfortunate circumstance in this case was that neglect by
the scientific community of Bessel's ingenious ideas slowed down the
progress in the understanding of the science behind the acceleration.

Nowadays a change from acceleration to deceleration (and vice versa) is nothing
exceptional among~the~short-period comets.  Some undergo such transitions over
a period of just several revolutions about the Sun.  Encke's comet was and still
is different in that the rate of the nongravitational acceleration in its orbital
motion was since 1820s fairly smoothly declining with time, subject to no major
erratic variations whatsoever.  The comet surely deserves this commentary! \\

This research was carried out at the Jet Propulsion Laboratory, California
Institute of Technology, under contract with the National Aeronautics and
Space Administration. \\

\begin{center}
{\footnotesize REFERENCES}
\end{center}

\vspace{-0.25cm}
\begin{description}
{\footnotesize
\item[\hspace{-0.3cm}]
Backlund, O. 1910, Mon. Not. Roy. Astron. Soc., 70, 429
\\[-0.57cm]
\item[\hspace{-0.3cm}]
Bessel, F. W. 1836a, Astron. Nachr., 13, 185 % recoil effect p. 232
\\[-0.57cm]
\item[\hspace{-0.3cm}]
Bessel, F. W. 1836b, Astron. Nachr., 13, 345 % critique resist. medium
\\[-0.57cm]
%
% Beyer, M. 1938, Astron. Nachr., 265, 37 %  Ho = 10.08 +/- 0.05 in 1937.99
% Beyer, M. 1950, Astron. Nachr., 278, 217 % Ho =  9.98 +/- 0.09 in 1947.90
% Beyer, M. 1955, Astron. Nachr., 282, 145 % Ho =  9.83 +/- 0.05 in 1951.20
% Beyer, M. 1962, Astron. Navhr., 286, 219 % Ho = 10.19 +/- 0.15 in 1961.10
%
\item[\hspace{-0.3cm}]
Beyer, M. 1972, Astron. Nachr., 293, 241 %   Ho = 10.83 +/- 0.10 in 1971.02
\\[-0.57cm]
\item[\hspace{-0.3cm}]
Encke, J. F. 1819, Berlin. Astron. Jahrb. f\"{u}r 1822, 180
\\[-0.57cm]
\item[\hspace{-0.3cm}]
Encke, J. F. 1820, Berlin. Astron. Jahrb. f\"{u}r 1823, 211
\\[-0.57cm]
\item[\hspace{-0.3cm}]
Encke, J. F. 1823, Berlin. Astron. Jahrb. f\"{u}r 1826, 124
\\[-0.57cm]
\item[\hspace{-0.3cm}]
Encke, J. F. 1859, Astron. Nachr., 51, 81 % Apps. 1855 & 1858
\\[-0.57cm]
\item[\hspace{-0.3cm}]
Jaegermann, R.\,1903, Prof.\,Dr.\,Th.\,Bredichin's Mechanische~Unter-{\linebreak}
 {\hspace*{-0.6cm}}suchungen \"{u}ber Cometenformen.  Leipzig, St.\,Petersburg:\
 Voss,{\linebreak}
 {\hspace*{-0.6cm}}500pp
\\[-0.57cm]
\item[\hspace{-0.3cm}]
Kam\'el, L. 1991, Icarus, 93, 226
\\[-0.57cm]
\item[\hspace{-0.3cm}]
Makover, S. G. 1955, Trudy Inst.\,Teor.\,Astron.\,Akad.\,Nauk~SSSR,{\linebreak}
 {\hspace*{-0.6cm}}4, 133
\\[-0.57cm]
\item[\hspace{-0.3cm}] 
Marsden, B. G. 1969, AJ, 74, 720
\\[-0.57cm]
\item[\hspace{-0.3cm}]
Marsden, B. G. 1970, AJ, 75, 75
\\[-0.57cm]
\item[\hspace{-0.3cm}] 
Marsden,\,B.\,G., \& Green,\,D.\,W.\,E.\,1985,~Quart.\,J.\,Roy.\,Astron.\,Soc.,{\linebreak}
 {\hspace*{-0.6cm}}26, 92
\\[-0.57cm]
\item[\hspace{-0.3cm}]
Marsden, B. G., \& Sekanina, Z. 1974, AJ, 79, 413
\\[-0.57cm]
\item[\hspace{-0.3cm}]
Marsden, B. G., Sekanina, Z., \& Yeomans, D. K. 1973, AJ,~78,~211
\\[-0.57cm]
\item[\hspace{-0.3cm}]
Meisel, D. D. 1969, Publ. Astron. Soc. Pacific, 81, 65
\\[-0.57cm]
\item[\hspace{-0.3cm}]
Nakano, S. 1985, Nakano Note NK\,485
\\[-0.57cm]
\item[\hspace{-0.3cm}]
Nakano, S. 1990, Nakano Note NK\,545
\\[-0.57cm]
\item[\hspace{-0.3cm}]
Nakano, S. 1997, Nakano Note NK\,658
\\[-0.57cm]
\item[\hspace{-0.3cm}]
Nakano, S. 2000, Nakano Note NK\,729
\\[-0.57cm]
\item[\hspace{-0.3cm}]
Nakano, S. 2004, Nakano Note NK\,1045
\\[-0.57cm]
\item[\hspace{-0.3cm}]
Nakano, S. 2005, Nakano Note NK\,1279
\\[-0.57cm]
\item[\hspace{-0.3cm}]
Nakano, S. 2007, Nakano Note NK\,1462
\\[-0.57cm]
\item[\hspace{-0.3cm}]
Nakano, S. 2010, Nakano Note NK\,1927
\\[-0.57cm]
\item[\hspace{-0.3cm}]
Nakano, S. 2013, Nakano Note NK\,2573
\\[-0.57cm]
\item[\hspace{-0.3cm}]
Nakano, S. 2017, Nakano Note NK\,3409
\\[-0.57cm]
\item[\hspace{-0.3cm}]
Nakano, S. 2020, Nakano Note NK\,4234
\\[-0.57cm]
\item[\hspace{-0.3cm}]
Rudenko, M. 2021, MPEC 2021-R75
\\[-0.58cm]
\item[\hspace{-0.3cm}]
Sekanina, Z. 1979, Icarus, 37, 420
\\[-0.58cm]
\item[\hspace{-0.3cm}]
Sekanina, Z. 1987, in Diversity and Similarity of Comets, ESA{\linebreak}
 {\hspace*{-0.6cm}}SP-278, ed.\ E. J. Rolfe \& B. Battrick.  Noordwijk,
 Netherlands:{\linebreak}
 {\hspace*{-0.6cm}}ESTEC, 315
\\[-0.57cm]
\item[\hspace{-0.3cm}]
Sekanina, Z. 1988a, AJ, 95, 911
\\[-0.57cm]
\item[\hspace{-0.3cm}]
Sekanina, Z. 1988b, AJ, 96, 1455
\\[-0.57cm]
\item[\hspace{-0.3cm}]
Sekanina, Z. 1991, J. Roy. Astron. Soc. Canada, 85, 324
\\[-0.57cm]
\item[\hspace{-0.3cm}]
Sekanina, Z. 1993, AJ, 105, 702
\\[-0.57cm]
\item[\hspace{-0.3cm}]
Whipple, F. L. 1950, ApJ, 111, 375
\\[-0.57cm]
\item[\hspace{-0.3cm}]
Whipple, F. L., \& Sekanina, Z. 1979, AJ, 84, 1894
\\[-0.63cm]
\item[\hspace{-0.3cm}]
Williams, G. V. 2017, MPEC 2017-L52}
% \\[-0.67cm]
\vspace{-0.39cm}
\end{description}
\end{document}